\documentclass[reprint,amsmath,amssymb,aip,pre,author-numerical]{revtex4-1}
\pdfoutput=1

\usepackage{graphicx}
\usepackage{hyperref} 
\usepackage{mathtools}
\usepackage{siunitx}
\usepackage{dcolumn}
\usepackage{amsmath}
\usepackage{amssymb}
\usepackage{xfrac}
\usepackage{bm}
\usepackage[dvipsnames]{xcolor}
\usepackage{tcolorbox}
\usepackage{float}

\newcommand{\dt}{\partial_\theta}

\DeclareMathOperator{\Li}{Li}
\newcommand{\laplace}[1]{\left[\Delta_{\mathcal{S}} + #1\right]}

\begin{document}
	
	\title{pH gradient-driven deformation of a crista-like vesicle}
	
	\author{Yorgos Chatziantoniou}
	\affiliation{UMR CNRS Gulliver 7083, ESPCI Paris, PSL Research University, 75005 Paris, France}%
	\author{Hélène Berthoumieux}
	\email{helene.berthoumieux@espci.psl.eu}
	\affiliation{UMR CNRS Gulliver 7083, ESPCI Paris, PSL Research University, 75005 Paris, France}%

	\begin{abstract}
		The inner membrane of mitochondria presents folds, the cristae, which are the production place of ATP. This synthesis is driven by a flow of protons confined to the surface of the membrane, which also shapes the crista to ensure a high synthesis rate. We model a crista as a spherical vesicle submitted to a diffusive proton gradient flowing from the poles to the equator. Using Helfrich model, we introduce a pH-dependent spontaneous curvature for the membrane and determine the shape of the vesicle, when submitted to the pH gradient, in the regime of small deformations. Based on biophysical arguments, we define a functionality score for the vesicle and construct a phase diagram identifying the zones of "well-functioning" cristae, which we compare to experimental measurements.
	\end{abstract}
	
	\maketitle
	
	\section{Introduction}
	Lipid membranes are an essential component of eukariotic cells, enclosing organelles and participating in important biochemical processes. The inner mitochondrial membrane is an excellent example of this double function. It forms dozens of invaginations called cristae. The shape of a crista can be tubular, ampular, or lamellar depending on the cell type,\cite{panek2020} but presents in a conserved manner flat and highly curved zones. A crista contains two types of transmembrane proteins, the respiratory complexes (RC) and dimers of ATP synthase (ATP-S), which are located in its flat and curved zones, respectively.~\cite{davies2011} RC pump protons from the matrix to the crista inner space. The membrane guides the protons,~\cite{serowy2003,knyazev,heberle1994,heberle2000,gennis2016} which diffuse in the interfacial water layer~\cite{cherepanov2003low} along the crista surface and flow through ATP-S, driving the ATP synthesis.\par 
	A three-dimensional representation of a crista is shown in Fig.~\ref{fig:1} (a). The crista membrane is represented by the white ellipsoid surface. RC are represented by a purple disk at the poles of the vesicle and ATP-S dimers by orange cones, which align along the equator of the vesicle. The blue arrows show the proton flux travelling from the RC to the ATP-S along the inner side of the membrane. The grey cylinder represents the crista junction that acts as a hub for ion regulation and molecule exchange between the interior of the crista and the rest of the mitochondrion.\par
	The crista membrane contains cardiolipins (CL), which are lipids found almost exclusively in these systems and are essential for the mitochondria to function efficiently.\cite{paradies2014,ikon2017} First, the conical shape of one lipid favors the formation of highly curved membranes,\cite{beltran2019} such as the edge of the crista. Second, cardiolipins are involved in an acid-base reaction at physiological pH as follows: HCL$^-\leftrightharpoons$CL$^{2-}$+H$^+$, which could make CL act as proton traps.\cite{heberle2000,haines2002,variyam24} 
	Finally, the mechanical properties of membranes containing CL are sensitive to pH. {\it In vitro} experiments on CL-containing giant unilamellar vesicles have shown that a localized pH gradient can remodel the membrane to form a crista-like invagination.\cite{khalifat2008membrane,khalifa2011} Recently, molecular dynamics simulations have confirmed that accumulation of protons along a CL-containing membrane modifies its spontaneous curvature.\cite{allolioc2021}\par
	The membrane of a crista is shaped by two players. Firstly, the proteins RC and ATP-S impose different constraints due to their geometry. Secondly, the protons injected into the system can influence the membrane composition by interacting with CL, thus modifying the mechanical properties of the membrane. Despite its importance in mitochondrial bioenergetics, a precise model linking the mechanical properties of membranes composed of CL, the H$^+$ flux, the protein constraints and the shape of an active crista membrane is lacking.\cite{joubert2021}
	
	Based on the minimization of the seminal Helfrich Hamiltonian,\cite{helfrich1973} the shapes of {\it in vitro} vesicles, their stability, transformations, and fluctuations have been extensively studied both experimentally and theoretically, despite cumbersome mathematics.\cite{helfrich1989,seifert1997} {\it In vivo} surfaces, such as the well-studied cell cortex, can also be described as Helfrich membranes whose mechanical properties are regulated by an active diffusing agent.\cite{mietke,thesemietke} The coupling between diffusion and mechanics give access to a new variety of shapes. Recently, the derivation of the energy-minimizing shape as a
	functional variation problem with geometric constraints has made the problem more accessible.\cite{Guv04a,Deserno_curvatureconvention} \par
	In this framework, the deformation of a pH-sensitive Helfrich membrane has been studied for flat and for cylindrical membranes.\cite{AF2,patil2020,mendes2023} In the second case, it has been shown that a diffusive proton flux can induce bulges and necks reminiscent of the {\it in vivo} cristae shape.\cite{patil2020,mendes2023} 
	Here we model the crista as a closed spherical vesicle that flattens under the effect of a proton current. This geometry mimics the crista shape observed by electron microscopy.\cite{davies2011,blum2019} The work is organized as follows. We model a crista at rest, with no synthesis of ATP and a vanishing proton flux, as a spherical vesicle.
	The surface proton concentration resulting from the activities of the RC located at the pole and the ATP-S dimers at the equator is governed by a purely diffusive process. 
	The mechanical properties of the crista membrane are modeled using the Helfrich Hamiltonian in which we introduce a spontaneous curvature depending on the local proton concentration. We derive the stress tensor induced by the 2D proton concentration for an active crista.\cite{patil2020,mendes2023} We determine the resulting shape of the vesicle imposed by the proton flux, (see sketch Fig.~\ref{fig:1}), in the regime of small deformations. The shape is controlled by two reduced parameters functions of the mechanical parameters of the Helfrich Hamiltonian. Finally, we define a functionality score for the vesicle, which measures how well the crista works based on biophysical arguments. We identify the parameter ranges for a "well-functioning" crista.

	\section{Proton flux as a driving force}

	\begin{figure*}
		\includegraphics[scale=0.30]{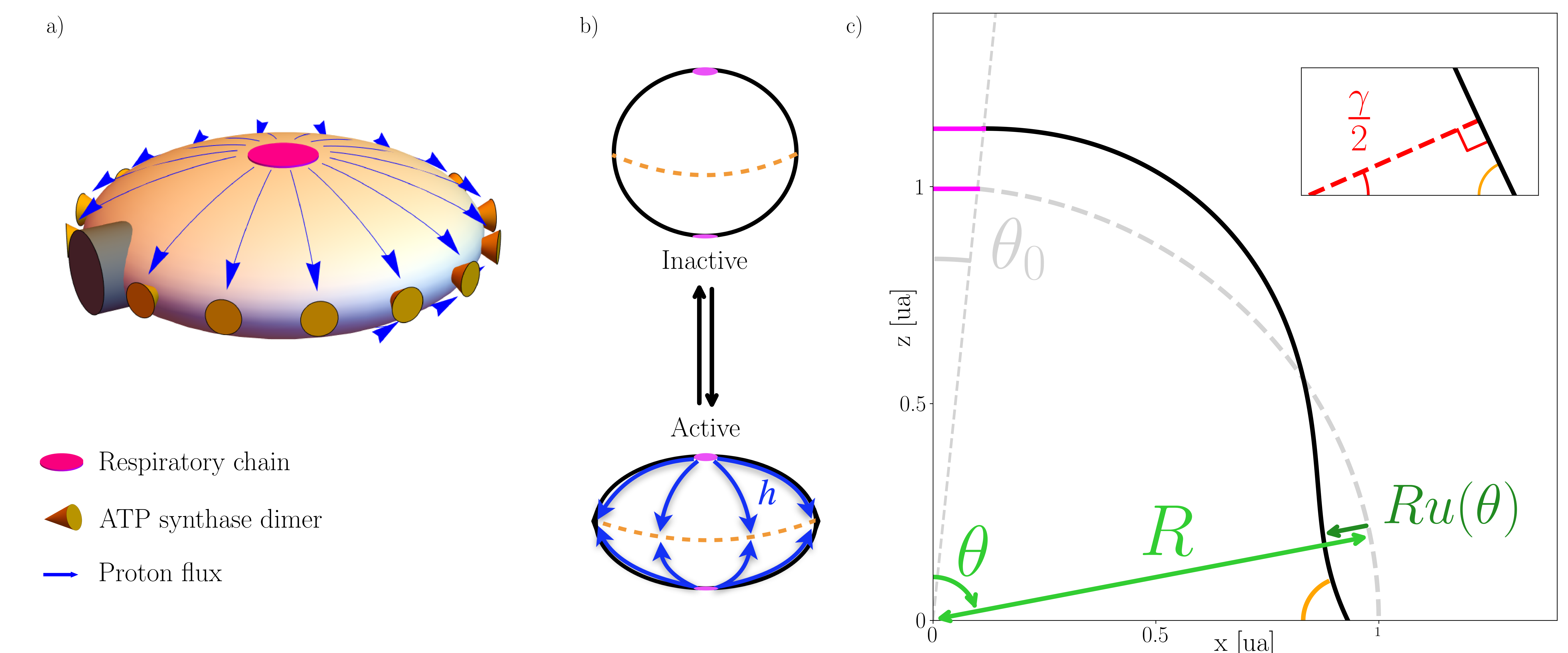} 
		\caption{Sketch and parameterization of the system. (a) Sketch of a functioning crista. The white ellipsoid represents the crista membrane, the purple patch the RC, the orange cones the ATP-S dimers and the blue arrows the surface proton flux. (b) The transition between a crista at rest, modeled as a spherical vesicle associated with a vanishing proton flux, and an active deformed crista submitted to a proton flux. (c) Generating curge of the crista at rest (dotted grey line) and of the active crista (black curve) in the (x,z) plane. The purple patch represents the RC as a disk of radius $r$=0.1$R$. The disk remains flat and constrains membrane slope at $\theta=\theta_0$, as shown in inset 1. The angle $\gamma$ is the opening angle of ATP-S dimers. It imposes membrane slope at $\theta=\pi/2$ as illustrated in inset 2.}
		\label{fig:1}
	\end{figure*}
\begin{figure}
		\centering
		\includegraphics[scale=0.3]{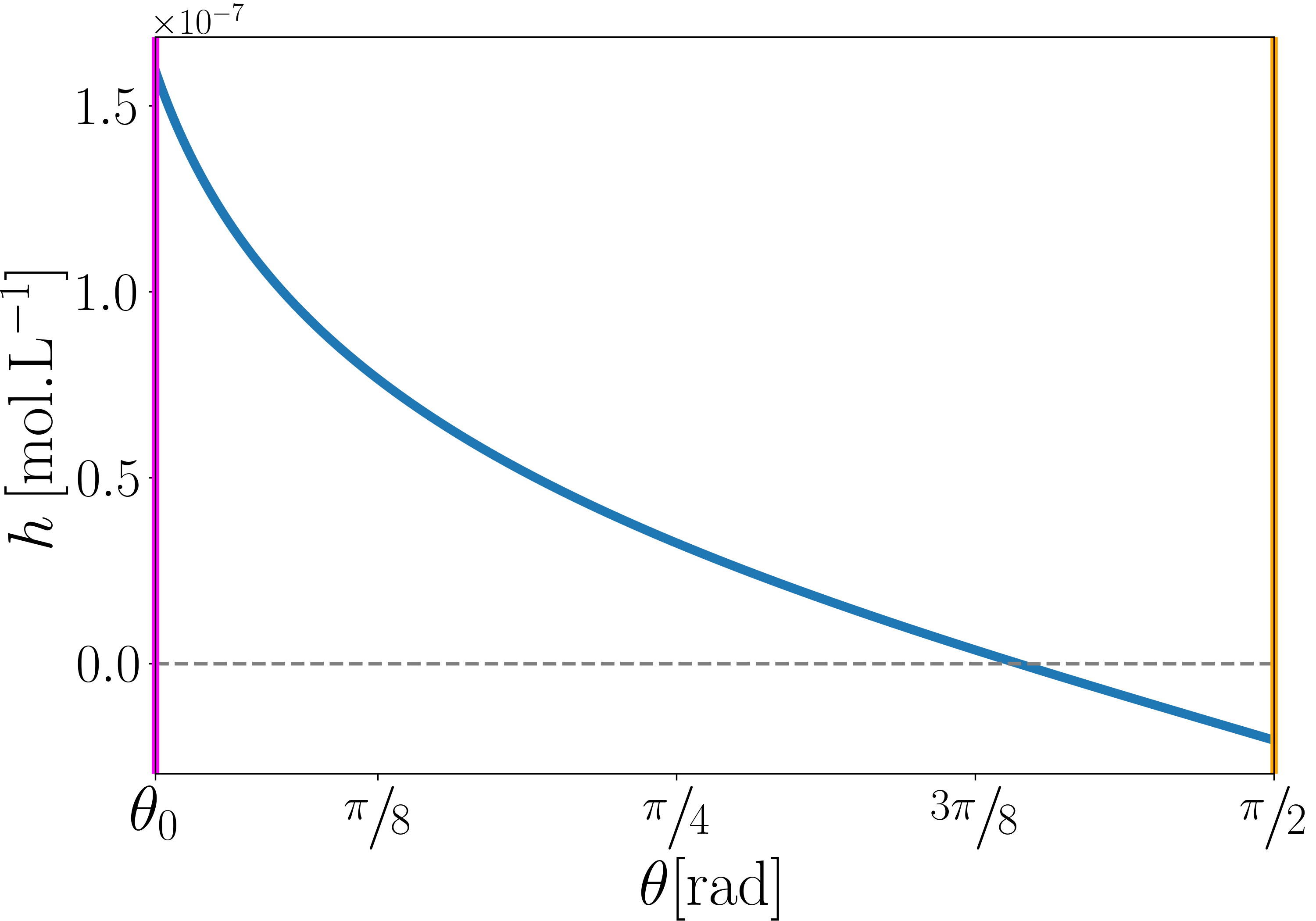} 
		\caption{Proton concentration along the membrane. We plot $h(\theta)$ given in Eq.~(\ref{proton_field}) for $\alpha_0$=$\qty{-6.0e-8}{\mole\per\liter}$ and $\alpha_1$ = $\qty{7.9e-8}{\mole\per\liter}$.\cite{rieger2014lateral} The purple and yellow vertical lines indicate the location of RC and ATP-S respectively.}
		\label{fig:2}
	\end{figure}

	We study a spherical system emulating a crista of radius \(R\), with an $xy$ planar symmetry and a $z$-axis rotational symmetry. It is thus parametrized by the polar angle \(\theta\). The surface of this pseudo-crista has two different phases: rigid patches at the north and south poles of radius $r$ represent the RC (in magenta in Fig~\ref{fig:1}) and a fluid membrane elsewhere. The boundary separating the two phases is located at the fixed angle \(\theta=\theta_0\approx r/R\).
	We model the proton concentration generated by RC and ATP-S activity, $h(\theta)$, as a purely 2D diffusive between RC ($\theta=\theta_0$) and ATP-S ($\theta=\pi/2$). $h(\theta)$ thus obeys: 
	\begin{eqnarray}
		\Delta_{\mathcal{S}} h(\theta)=0, \quad \theta \in [\theta_0,\pi/2]
	\end{eqnarray}
	where $\Delta_{\mathcal{S}}=\left(\partial_\theta+{\rm cotan}\!\left(\theta\right)\right)\dt$ is the 2D spherical Laplacian operator. We obtain
	\begin{align}
		\label{proton_field}
		h(\theta)=& - \alpha_0 \ln \left(\tan \left(\frac{\theta}{2}\right)\right)+\alpha_1, \quad \alpha_0,\alpha_1\in\mathbb R
	\end{align}
	We define the pH in the water layer hydrating the membrane as follows: pH($\theta$)=-log ($h_0+h(\theta)$), where $h_0$=$\qty{10e-7}{\mole\per\liter}$ is the physiological bulk proton concentration in the crista in the absence of ATP production. The pH in the cristae has been measured experimentally using {\it in situ} fluorescence spectroscopy and estimated to pH = 6.4 close to RC and pH = 7.1 close to ATP-S.\cite{rieger2014lateral} We fit $\alpha_0, \alpha_1$ to these data. Their values are given in the caption of Fig.~\ref{fig:2}. Figure~\ref{fig:2} shows the concentration of protons, $h(\theta)$, and
	we see that it decreases monotonically with $\theta$ from the location of RC (purple line) to the equatorial edge and the ATP-S location (yellow line). $h(\theta)$ is positive over almost the whole surface, indicating an accumulation of protons with respect to the physiological concentration, but becomes negative close to ATP-S, indicating a lack of protons in this zone.
	
	\section{pH-dependent Helfrich Hamiltonian}
	The surface \(\mathcal S\) of the membrane, enclosing a volume \(\mathcal V\), is described by a 3D vector $\mathbf{X}$, which is parametrized by the two angles $\theta$ and $\phi$. At rest, the crista is spherical, and $\mathbf X = \mathbf X_0$ :
	\[
	\begin{matrix}
		\mathbf X_0 : & [\theta_0, \sfrac\pi2] \times \left[\right.0,2\pi\left[\right. & \longrightarrow& \mathbb R^3\\
		&(\theta, \phi)&\longmapsto& \begin{pmatrix}R\sin\theta \cos\phi\\ R\sin\theta \sin\phi\\ R\cos\theta\\\end{pmatrix}.
	\end{matrix}
	\]
	When the crista is active, its surface is deformed by a perturbative field $\mathbf X_1$, that we assume to be radial :
	
	\[
	\begin{matrix}
		\mathbf X_1 : &[\theta_0, \sfrac\pi2] \times \left[\right.0,2\pi\left[\right. & \longrightarrow &\mathbb R^3\\
		&(\theta, \phi) & \longmapsto &Ru(\theta) \begin{pmatrix}\sin\theta\cos\phi \\ \sin\theta\sin\phi\\ \cos\theta \\\end{pmatrix},
	\end{matrix}
	\]
	with \(u(\theta)\) the dimensionless axisymmetric deformation field, illustrated in Fig.~\ref{fig:1}c. In the spherical basis, $(\hat{\mathbf x}_r, \hat{\mathbf x}_\theta, \hat{\mathbf x}_\phi)$, the surface vector can be written as
	\begin{equation}
		\label{deformed_sphere}
		\mathbf X (\theta,\phi) = \mathbf X_0(\theta, \phi) + \mathbf X_1(\theta,\phi) = R(1 + u(\theta)) \hat{\mathbf x}_r(\theta,\phi), 
	\end{equation}
	We focus on the small deformation regime, $u(\theta)$ obeys $u(\theta)\ll 1$.
	From Eq.~(\ref{deformed_sphere}), one can then obtain the local intrinsic basis of $\bf {X}$, $({\bf e}_\theta, {\bf e}_\phi)$, defined as ${\bf e}_a =\partial_a {\bf X}$, $(a=\theta, \phi)$, and the normal vector ${\bf n}={\bf e}_\theta \times {\bf e}_\phi/| {\bf e}_\theta \times {\bf e}_\phi|$, 
	a vector of unit length perpendicular to the surface. We can furthermore define the two fundamental forms of the surface, i.e. the curvature tensor $K_{ab}={\bf e}_a \cdot \partial_b {\bf n}$
	 and the metric tensor $g_{ab}={\bf e}_a \cdot {\bf e}_b$. Note that the twice covariant metric tensor, \(g^{ab}\), is such that: $g^{ab}=g_{ab}^{-1}$. The surface element $d^2\mathbf X$ is equal to: $d^2\mathbf X=\sqrt{g} d\theta d\phi$, where $g =\det g_{ab}$ 
	is the determinant of the metric tensor. For the tensorial calculations below, 
	we remind the Einstein summation convention and the passage from covariant to contravariant coordinates, such that $T^a_b=T_{bk}g^{ka} = \sum_{k=1,2} T_{bk}g^{ka}$ for ${\bf T}$ a tensor of rank 2.\par
	To model the mechanical properties of the membrane of the crista, we use the Helfrich model for elastic membranes, developed by
	Wolfgang Helfrich in 1973.\cite{helfrich1973} This effective energy functional of a membrane takes into account molecular properties of lipid membranes such as the fluidity and the absence of
	in-plane shear stress but is written at a continuous coarse-grained scale. We modify the standard Helfrich model by introducing a pH-dependent spontaneous curvature as:
	
	\begin{equation}
		\label{pH-helfrich}
		H[\mathbf{X},h] = \int_{\mathcal S}\mathrm d^2\mathbf X \left[\frac12\kappa\left(C-C_0\left[h\right]\right)^2 + \sigma_0\right] - P\mathcal V
	\end{equation}
	The parameter $\kappa$ is the bending modulus, $\sigma_0$ is the surface tension and
	$C$ is the local curvature. The Gaussian bending rigidity has been neglected for the sake of simplicity. As we model the crista as a closed system, we introduce the pressure $P$ as a Lagrange multiplier to maintain constant volume. \par The spontaneous curvature $C_0(h)$ is assumed to couple to the proton concentration in a linear perturbative manner as follows:
	\begin{equation}
		\label{C0pH}
		C_0(h) = C_0^0 + C^1_0 h(\theta),
	\end{equation}
	with $C_0^0$ the spontaneous curvature of the membrane at pH=7 and $C^1_0$ its sensitivity to \(h\). We consider a small perturbation so we set $C_0^1$ such that $|C_0^1 h(\theta)|\ll 1/R$. Note that \(C_0(\theta)\) is not constant but varies along the surface.\par

	The stationary shape of the deformed crista satisfies the shape equation, \textit{i.e.} the normal force balance on the membrane, derived from the stress tensor \(\mathbf f\). Following the approach of Guven and coworkers ,\cite{Guv04a,Muller_phd,Deserno_curvatureconvention} we expand the Hamiltonian in Eq.~(\ref{pH-helfrich}) by introducing the geometrical characteristics of the surface: the instrinsic base, the metric and curvature tensors and the normal vector. Instead of using the relations that bind them together, we consider them as free fields that we then constrain with Lagrange multipliers. Finally, we recognize the multiplier constraining the vectors of the basis, \(\mathbf f^{a}\), as the stress tensor.
	Note that the membrane is a 2D space that lives in \(\mathbb R^3\), thus the tensor\(f\) can be described as a \(3\times2\) matrix.\cite{fournier2007} The generic expression for $\mathbf{f}$ is~\cite{Guv04a,Muller_phd,Deserno_curvatureconvention}
	
	\begin{widetext}
		\begin{align}
			\mathbf f^a &= \left[-\left(\frac12\kappa\left(C-C_0\right)^2 + \sigma_0\right)g^{ab} + \kappa \left(C-C_0\right)K^{ab}\right] \mathbf e_b + \nabla^a\left(C-C_0\right)\mathbf{n},
			\label{stresstensor}
		\end{align}
	\end{widetext}
	with $\mathbf{f}^{a}= f^{ab}\mathbf{e}_b+f^{an}\mathbf{n}$, $(a,b)\in(\theta,\phi)$. The contributions $f^{ab}$ and $f^{an}$ are the transverse and normal components of the stress tensor. $\nabla_a$ is the covariant derivative.
	
	With these definitions, the force balance on the surface $\mathbf{X}$ reads
	\begin{align}
		\label{tangent-force-balance}
		\nabla_a\mathbf{f}^{ab}-K_a^bf^{an}&=0\\
		\label{normal_force_diffgeom}
		\nabla_a\mathbf{f}^{an}+K_{ab}f^{ab}&=P.	
	\end{align}
	Eq.~(\ref{tangent-force-balance}) corresponds to the force balance along the tangential directions and Eq~(\ref{normal_force_diffgeom}) to that along the normal direction. Due to the symmetries of the problem, the latter is the only equation that contains useful information. Using Eq.~(\ref{stresstensor0}), it can be written as,
	\begin{widetext}
		\begin{align}
			\label{general_shape_equation}
			\left[\frac12\kappa\left(C-C_0\right)^2 + \sigma_0\right]C - \left(C-C_0\right)\left(C^2 - 2K_G\right) -\kappa \Delta\left(C-C_0\right) = P.
		\end{align}
	\end{widetext}
	
	$K_G$ is the Gaussian curvature of the membrane, defined as the determinant of the curvature matrix $K_a^b$ and $\Delta$ is the Laplace-Beltrami operator of the surface $\mathcal{S}$. 
	In the deformed sphere, Eq.~(\ref{deformed_sphere}), the stress tensor $\mathbf{f}$ can be expanded as follows
	\begin{equation}
		{\bf f}={\bf f}^0+\delta {\bf f},
	\end{equation}
	with ${\mathbf f}^0$, associated with the perfect sphere, and $\delta {\mathbf f}$ the first-order contribution in the deformation field $u$ and the driving force field $h$. 
Using expressions given in Appendix \ref{ap:geodiff0}, we derive
	\begin{widetext}
		\begin{align}
			\label{stresstensor0}
			\mathbf{f}^0&=\begin{pmatrix}
				\kappa C_0^0R^{-3}-\frac12\kappa (C_0^0)^2R^{-2} - R^{-2}\sigma_0 & 0\\
				0 &\sin^{-2}\theta\left[\kappa C_0^0R^{-3}-\frac12\kappa (C_0^0)^2R^{-2} - R^{-2}\sigma_0\right]\\
				0 & 0 
			\end{pmatrix},
		\end{align}
	\end{widetext}
	which is purely tangential, as $f^{an}$ vanishes, and~\ref{ap:geodiff1}
	\begin{widetext}
		\begin{align}
			\label{stresstensor1theta}
			\delta_{\rm mech} \mathbf{f}^\theta&=\kappa R^{-4}\begin{pmatrix}
				\cot\theta u' - u'' - C_0^0R\left(3u+\cot\theta u'\right) + \left(C_0^0R\right)^2 u + 2 R^2\sigma_0u\\
				0\\
				\left[\left(1-\cot^2\theta\right)u' +\cot\theta u'' + u^{(3)} \right]R
			\end{pmatrix}\\	
			\label{stresstensor1thetachem}
			\delta_{\rm chem} \mathbf{f}^\theta&=\kappa R^{-4}\begin{pmatrix}
				\left(1-RC_0^0\right)RC_0^1h\\
				0\\
				R^2 C_0^1 h'
			\end{pmatrix}\\	
			\label{stresstensor1phi}
			\delta_{\rm mech} \mathbf{f}^\phi&=\kappa R^{-4}\sin^{-2}\theta\begin{pmatrix}
				0 \\
				u'' - \cot\theta u' - C_0^0R\left(3u+u''\right) +\left(C_0^0R\right)^2u +2R^2\sigma_0u \\
				0 
			\end{pmatrix}\\	
			\label{stresstensor1phichem}
			\delta_{\rm chem} \mathbf{f}^\phi&=\kappa R^{-4}\sin^{-2}\theta\begin{pmatrix}
				0 \\
				\left(1-RC_0^0\right) RC_0^1h\ \\
				0 
			\end{pmatrix}
		\end{align}
	\end{widetext}
	where the superscript $'$ denotes the spacial derivative with respect to $\theta$. Note that the apparent inhomogeneity in physical dimensions between $f^{\theta\theta}$ and \( f^{\theta n}\) is due to the definition of the vectors \(\mathbf e_i\), that have the dimension of a length, and \(\mathbf n\), a dimensionless direction.
	The normal first-order contribution of the stress tensor, $\delta \mathbf{f}^{\theta n}$ is a function of $h(\theta)$ ; the variation of $h$ along the surface induces active tensions and bending which drive the deformation of the crista membrane.
	
	Applied to a perfect sphere, Eq~(\ref{general_shape_equation}) leads to the shape equation that was first derived by Helfrich:~\cite{helfrich1987,helfrich1989,seifert1997}
	\begin{equation}
		\label{shape_equation}
		\kappa (C_0R)^2 - 2\kappa C_0R + 2R^2\sigma_0 = R^3 P.
	\end{equation}
	\par
		Before extending the study to the deformed sphere, it is useful to reduce the number of parameters in the problem by introducing dimensionless parameters.
\begin{tcolorbox}[colback=JungleGreen!5, colframe=JungleGreen!40!black, title=A nice heading, fonttitle=\bfseries, title=Introduction of dimensionless parameters]
 We choose the radius $R$ of the sphere and the bending elasticity \(\kappa\) of the membrane to normalize the three other parameters $(P, \sigma_0, C_0)$ of the system and give the expression of the associated dimensionless parameters $(p, \sigma, \xi)$ in Tab.~\ref{dimensionless_parameter}. $\kappa$ and \(R\) become effectively the units of the problem. Using Eq.~(\ref{shape_equation}), we can express one of the three remaining parameters ($p$, $\sigma$, $\xi$) as a function of the last two, reducing the number of independent parameters from 5 to 2. This reduction is mathematical only; physically, the values of \(\kappa\) and \(R\) still matter.
	\begin{table}[H]
		\centering
		\begin{tabular}{c|c|c}
			Physical quantity & Dimensionless quantity & Definition \\
			\hline
			\(P\) & \(p\) & \( R^3\kappa^{-1}P\)\\
			\hline
			\(\sigma_0\) & \(\sigma\) & \( R^2\kappa^{-1}\sigma_0\)\\
			\hline
			\(C_0\) & \(\xi\) & \( RC_0\)\\
			\hline
			\hline
			\(C_0^1h(\theta)\) & \(\delta\xi(\theta)\) & \( R\-C_0^1 h(\theta)\)\\
			
		\end{tabular}
		\caption{Dimensionless quantities characterizing the system. The physical parameters $P$, $\sigma_0$ and $C_0$ of the membrane and the active field $C_0^1$  have been normalized using $R$ and $\kappa$.}
			\label{dimensionless_parameter}
	\end{table}
	
\end{tcolorbox}

	We derive the normal force balance Eq.~(\ref{normal_force_diffgeom}) for the deformed shape. To do this, we inject the first-order expression of the curvature tensor $K_{ab}$ and of the surface gradient $\nabla_a$ (see Appendix~\ref{ap:geodiff1}) in the shape equation Eq.~(\ref{general_shape_equation}), and expand the latter at the first order in $u$ and $h$. This results in a non-homogeneous 4th-order differential equation in $u$, which, in dimensionless form, is as follows,
	
	\begin{equation}
		\label{deformation}
		\left(\Delta_{\mathcal S} + 2\right) \left(\Delta_{\mathcal S} + \mathcal A\right) u(\theta) = -\left[\Delta_{\mathcal S} + 2\left(\xi - 1\right)\right] RC_0^1h(\theta) + \delta p, \quad 
	\end{equation}
	with $\delta p$ the variation of the inside pressure induced by the deformation and with
	\begin{equation}
		\mathcal A = \xi - \frac{p}{2}.
	\end{equation}
	Eq.~(\ref{deformation}) exhibits a new parameter, \(\mathcal A\), that together with \(\xi\), forms a handy basis of parameters for the exploration of the phase space deformations of the membrane. In this basis, we find \(p = 2(\xi -\mathcal{A})\) and \(\sigma = 2\xi - \sfrac12\xi^2 - \mathcal A\).\par
	The force driving the deformation obeys
	\begin{align}
		\delta\xi (\theta) &= \beta h(\theta). 
	\end{align}
	where the coupling parameter $\beta$ is defined as ${\beta = RC_0^1}$. We choose \(\beta > 0\), which means that an excess of protons induces an increase in spontaneous curvature and a surface more curved towards the inside of the crista, in agreement with experimental and molecular dynamics simulation data.\cite{khalifat2008membrane,konar2023} We set the value of $\beta$ such that the amplitude of variation of the natural curvature is of the order of 10\% of the curvature of the system. Its value is given in the caption of Fig.~\ref{fig:3}. \par
	For a vanishing proton flux, $h(\theta)=0$, Eq.~(\ref{deformation}) leads to:
	\begin{align}
		\label{passive_crista}
		\left(\Delta_{\mathcal S} + 2\right) \left(\Delta_{\mathcal S} + \mathcal A\right) u =\delta p\,
	\end{align}
	which we identify as the equation governing the stability of a spherical membrane derived by Helfrich and Zhong-can using the second variation of the Helfrich Hamiltonian.\cite{helfrich1989} 
	Noting that $\Delta_{\mathcal S}h=0$, and of course \(\Delta_{\mathcal S} \delta p = 0\), we can rewrite Eq.~(\ref{deformation}) for \(\mathcal A \neq 0\) as follows,
	\begin{align}
		\label{homogeneous_deformation}
		\left(\Delta_{\mathcal S} + 2\right) \left(\Delta_{\mathcal S} + \mathcal A\right) \left[u - \frac{1-\xi}{\mathcal A} \beta h - \frac{1}{2\mathcal A} \delta p\right] &= 0
	\end{align}
	In other words, 
	\begin{equation}
		\label{part_sol}
		 u_{\rm p} (\theta) = \frac{1-\xi}{\mathcal A} \beta h(\theta) + \frac{1}{2\mathcal A}\delta p
	\end{equation}
	is a particular solution of Eq.~(\ref{deformation}). See Appendix~\ref{ap:a=0} for \(\mathcal A = 0\). \par Now, we need to solve the homogeneous equation, Eq.~\ref{passive_crista} where \(\delta p=0\), which amounts to solving the kernel of the operator 	$\left(\Delta_{\mathcal S} + 2\right) \left(\Delta_{\mathcal S} + \mathcal A\right)$.
	 For any \(\eta \in \mathbb R\), the functions
	\begin{widetext}
		\begin{eqnarray}
		\label{GF1}
		C_{\eta,0}(\theta) = \sum_{n\geqslant0}\prod_{k = 0}^{n-1}\left[2k\left(2k+1\right)-\eta\right] \frac{\cos^{2n}(\theta)} {\left(2n\right)\!!},\quad
		C_{\eta,1}(\theta) = \sum_{n\geqslant0}\prod_{k = 1}^{n}\left[\left(2k-1\right)2k-\eta\right] \frac{\cos^{2n+1}(\theta)} {\left(2n+1\right)\!!}\,
		\end{eqnarray}
	\end{widetext}
	provide a basis for the kernel of $\left(\Delta_{\mathcal S} + \eta\right)$. Note that the subscripts (0,1) refer to even and odd  functions respectively, relative to $\pi/2$.	
In particular for $\eta=2$, there are simpler expressions:
	\begin{eqnarray}
		\label{GF3}
		C_{2,0} (\theta) = \cos(\theta)\ln(\tan(\frac{\theta}{2}))+1, \quad
		C_{2,1} (\theta) = \cos(\theta).	
	\end{eqnarray}
For $\mathcal{A}\neq 2$, the functions $(C_{\mathcal{A},0}, C_{\mathcal{A},0}, C_{2,0} ,C_{2,0} )$ provide the basis of the kernel we are looking for. 	
	The degenerated case for \(\mathcal A = 2\) is solved in Appendix~\ref{ap:a=2}.
	Figure~\ref{fig:vecteurs_propres} shows a 3D representation of the generating functions \(C_{\mathcal A,0}(\theta) \text{ and } C_{\mathcal A,1}(\theta) \) for specific values of \(\mathcal A\). Note that \(C_{\eta,i}(\theta) \) is an infinite series and we approximate to its $n^{\rm th}$ truncation defined in Appendix \ref{ap:truncation}.
	\begin{center}
		\begin{figure*}
			\centerline{\includegraphics[scale=.38]{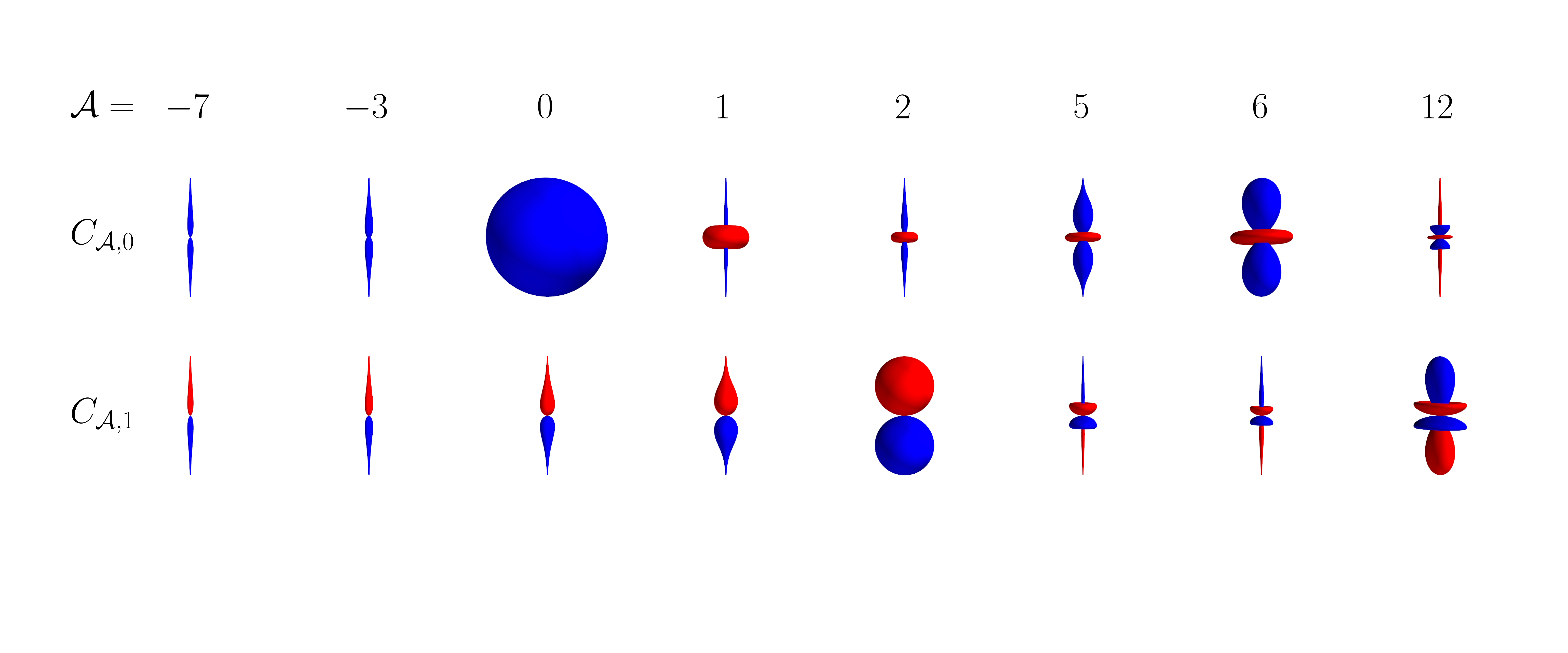} }
			\caption{Generating functions of $\left(\Delta_{\mathcal S} + \mathcal{A}\right)$. We show a 3D representation of \(C_{\mathcal{A},0}(\theta) \text{and} C_{\mathcal{A},1}(\theta) \) given in Eq.~(\ref{GF1}).  Blue/Red denotes a change of sign. For $\mathcal{A}=$0, 2, 6, 12, we recognize the four first spherical harmonics, truncated at the poles.}
			\label{fig:vecteurs_propres}
		\end{figure*}
	\end{center}
	
	Moreover, the particular solution $ u_{\rm p}(\theta)$, Eq.~(\ref{part_sol}), belongs to the kernel of $\Delta_\mathcal{S}$, as $\Delta_\mathcal{S} u_{\rm p}(\theta)=0$, and the functions
	\begin{equation}
		C_{0,0}(\theta) = 1, \quad C_{0,1}(\theta) = -\ln(\tan(\sfrac\theta2)
	\end{equation}
	provide a base for it.
	With all of this in mind, we can finally write the general form for $u(\theta)$ as follows:
	\begin{equation}
		\label{express_u}
		u(\theta) = \sum_{\substack{a = \mathcal A, 0, 2\\b = 0,1}} \lambda_{a,b}C_{a,b}(\theta). 
	\end{equation}
	We identify $\lambda_{0,0} = \beta(1-\xi)/\mathcal{A}\alpha_1 + \delta p/{2\mathcal A}$, and $\lambda_{0,1} = \beta(1-\xi)/\mathcal{A}\alpha_0$, using Eq.~(\ref{part_sol}) and Eq.~(\ref{proton_field}).\par
	To determine $(\lambda_{a,b})$, with $a \in [\mathcal{A},2]$, $b\in [0,1]$, we first set 4 boundary conditions that reflect the biological requirements for the crista to be able to function.\par
	First, at the site of insertion of the proteins, for \(\theta = \theta_0\), and  \(\theta =\sfrac\pi2\), the slope of the membrane must be adapted to the protein shape. At the pole, the RC are flat and force a flat membrane around them. The generative curve of the deformed surface is parallel to the $x$-axis, at $\theta=\theta_0$, as shown in Fig.~\ref{fig:1}. It gives 
	\begin{align}
		\label{membrane_slope_respiratory}
		-\frac rR u(\theta_0) + \cos(\theta_0)u'(\theta_0)&=\frac rR\ .
	\end{align}
	At the equator, the ATP-S dimers, that have an opening angle of $\gamma$ between the two monomers, impose an angle $(\pi-\gamma)/2$ between the membrane and the $x$-axis, in $\theta=\pi/2$, as illustrated in Fig.~\ref{fig:1}. This gives the following condition for the deformation field:
	\begin{align}
		\label{membrane_slope_ATP}
		\tan(\sfrac{\gamma}{2})u'(\sfrac\pi2) - u(\sfrac\pi2)&=1.
	\end{align} 
	Second, the proteins counter the force exerted by the membrane. At the pole,  we model the RC patch as a stiff spring with a surface tension of \(\Gamma_0\). The tangential force balance between the membrane and the spring can be written as follows,
	\begin{align}
		\delta_{\rm chem} f^{\theta\theta}(\theta_0)+	\delta_{\rm mech} f^{\theta\theta}(\theta_0)&= \Gamma_0\sin(\theta_0) u(\theta_0).
	\end{align}
At the equator, we model the force exerted by the ATP dimer on the membrane as a bending force with an elasticity constant $k$. This leads to the following equation:
	\begin{align}
		\label{boundary_ATP}
		\delta_{\rm chem} f^{\theta\theta}(\pi/2)+	\delta_{\rm mech} f^{\theta\theta}(\pi/2)&=k\cos(\gamma/2)u(\pi/2)
	\end{align}
	
	Note that the two latter equations are not dimensionless: they explicitly depend on \(\kappa\) and \(R\). We choose their values to be consistent with features of biological membranes, they are given in the caption of Fig.~\ref{fig:3}. 
	
	In addition to the boundary conditions, we adjust the pressure variation $\delta p$ so that the volume variation $\delta V$ vanishes:
	\begin{equation}
		\label{volume_variation}
		\delta V=	\int \!\sin(\theta)\textrm d\theta \,u(\theta) = 0.
	\end{equation}
	We use the expression for $u(\theta)$ given in Eq.~(\ref{express_u}) and inject it in Eqs.~(\ref{membrane_slope_ATP}-\ref{boundary_ATP}) to obtain linear equations for the coefficients $\lambda_{a,b}$, with $a \in [\mathcal{A},2]$, $b\in [0,1]$.
	The linearity of Eqs.~(\ref{membrane_slope_respiratory}-\ref{volume_variation})
	allows us to express the boundary conditions in the following matrix form:
	\begin{equation} 
		\label{lambda_solve}
		\mathbf M \cdot \mathbf \Lambda = \mathbf V,
	\end{equation}
 While there are 5 boundary conditions fixing $\left(\lambda_{\mathcal{A},0},\lambda_{\mathcal{A},1},\lambda_{2,0},\lambda_{2,1},\lambda_{0,0}\right)$, it is convinient to introduce a 6$^{\rm th}$ equation to fix the coefficient $\lambda_{0,1}$ such as 
	\begin{equation} 
\lambda_{0,1}=\alpha_0\beta\frac{1-\xi}{\mathcal{A}}.
	\end{equation}
	Thus  $ \mathbf M$ is a 6x6 matrix given in appendix~\ref{ap:matrix} and $\Lambda=\left(\lambda_{\mathcal{A},0},\lambda_{\mathcal{A},1},\lambda_{2,0},\lambda_{2,1},\lambda_{0,0},\lambda_{0,1}\right)$. There is no singularity in the phase space. We can thus always resolve Eq.~(\ref{lambda_solve}) and the corresponding deformation field $u(\theta)$, using Eq.~(\ref{express_u}). \par
	We compute $u(\theta)$ for different values of $(\mathcal{A}, \xi)$. We plot the parametric generative curve $(x(\theta)=R(1+u(\theta))\sin(\theta), z(\theta)=R(1+u(\theta))\cos(\theta))$ of the corresponding deformed spheres in Fig.~\ref{fig:3}. The four panels of Fig.~\ref{fig:3} show plots for increasing values of $\mathcal{A}$ from the left ($\mathcal{A}$=-7) to the right ( $\mathcal{A}$=5) and each panel shows curves for increasing values of $\xi$, from $\xi=-10$ (light blue curve) to $\xi$=10 (fushia curve).
	\begin{figure*}
		\includegraphics[scale=0.65]{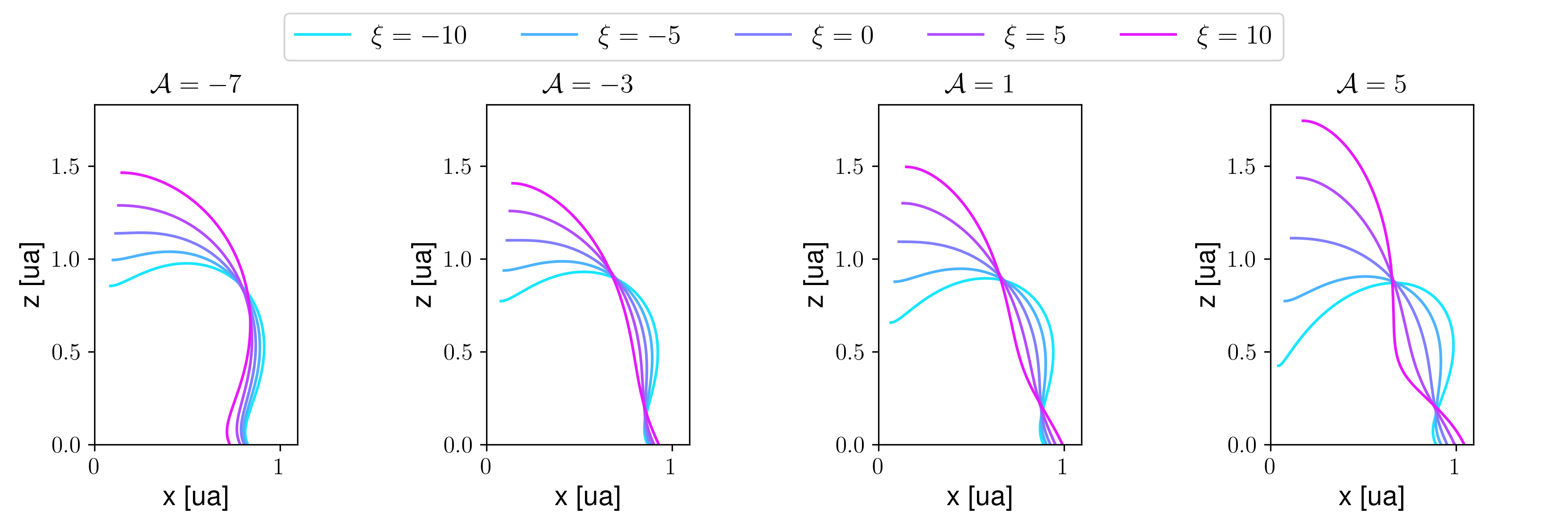} 
		\caption{Profiles of the deformed crista as a function of ($\mathcal{A}$, $\xi$). $u(\theta)$ obeys Eq.~(\ref{express_u}) and we solve Eq.~(\ref{lambda_solve}) for $\Gamma_0$=100\(\sigma_{\rm typ}\), with the typical surface tension of a bilipidic membrane \(\sigma_{\rm typ} = 1.10^{-7}\si[inter-unit-product = .]{\joule\per\square\meter}\)\cite{AF2} and $k = 100\kappa$. In our study, we set \(R = 1.10^{-7}\si{\meter}\) and \(\kappa = 6.10^{-20}\si{\joule}\).
		The other parameter values are given in caption of Figs.~(\ref{fig:1},\ref{fig:2})}
		\label{fig:3}
	\end{figure*}
	For a given $\mathcal{A}$ we observe a transition as a function of $\xi$ from oblate shape ($\xi \ll 0$) to prolate shape ($\xi \gg 0$). For $\xi$ given, a decreasing value of $\mathcal{A}$, corresponding to a higher value of $p$, leads to a more tense, less deformable sphere, associated with small values of the deformation field. On the contrary, high values of $\mathcal{A}$ correspond to a floppy shape, with a larger deformation field, especially for large $|\xi|$ (see the panel for $\mathcal{A}=5$).
	
	\section{Classification of the shapes}
	To discuss the shapes we obtain and their compatibility with the biological function of a crista, we introduce a functionality score $S$. This score is based on three geometrical key points.
	First we compute $S_1$, the perimeter of the equator of the crista, renormalised by the circumference of the sphere.
	Second, we compute $S_2$, the length of the path for the protons from RC to ATP-S, renormalised by the length of the sphere. Finally, we quantify $S_3$, the "flatness" in the region around RC, by calculating the absolute value of the mean curvature from $\theta=\theta_0$ to $\theta=\pi/16$, renormalised to the value for a perfect sphere. The three contributions can be written as follows.
	\begin{align}
		S_1&=\frac{2\pi R(1+u(\pi/2))}{2\pi R}, \nonumber \\ S_2&=\frac{l_H}{(\pi/2-\theta) R},\nonumber \\ S_3&=\frac{\int^{\pi/16}_{\theta_0} d\theta |C(\theta)|}{\pi/16} \nonumber
	\end{align} 
	The functionality score is defined as
	\begin{align}
		\label{score}
		S=S_1-S_2-S_3
	\end{align} 
	We compute the deformation field $u(\theta)$ for $\xi, \mathcal A \in [-10,10]$
	and calculate the corresponding score for each shape obtained. The color map  for $S$ as a function of $\xi$ and $\mathcal{A}$ is shown in Figure 4. A high value for $S$ (yellow regions on the density map) corresponds to a well-functioning crista, as it will be associated with a large equator and thus a high  number of ATP-S dimers (high $S_1$), a short path implies a better coupling between RC and ATP-S (small $S_2$),\cite{sjoholm2017} and a flat membrane around RC (small $S_3$) ensuring well-functioning proteins. On the contrary, the red regions, which correspond to low values of $S$ are associated with badly-functioning cristae. We see that well-functioning cristae lie in a valley around $\xi=-1$, which is deep and narrow for large values of $\mathcal{A}$, and widens and flattens as $\mathcal{A}$ decreases. 
	\begin{figure*}
		\centering
		\includegraphics[scale=0.5]{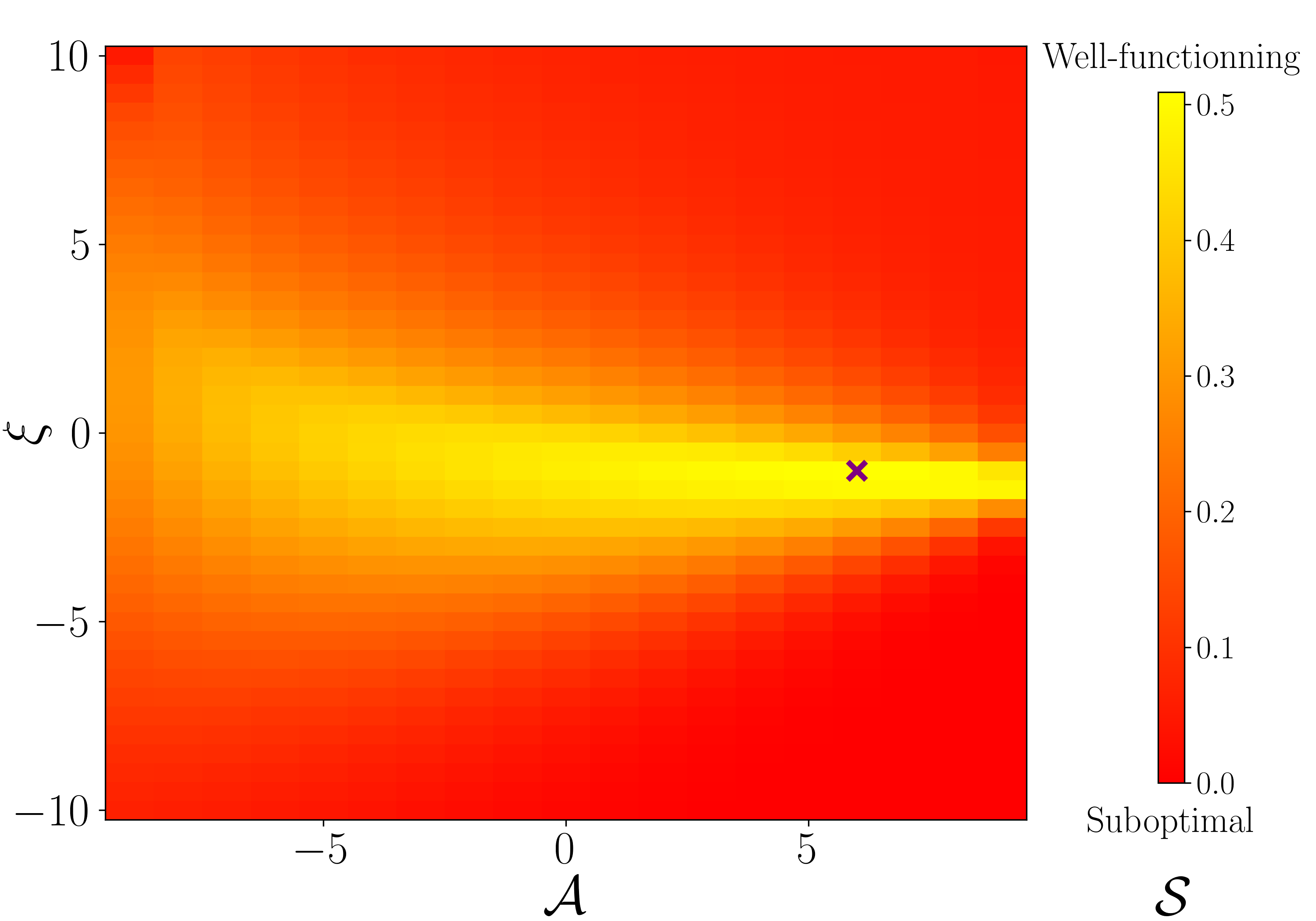}
		\caption{Functionality score for the crista. We represent the functionality score defined in Eq.~(\ref{score}) as a density map, function of $\mathcal{A}$ and $\xi$. The point of optimal score is marked with a purple cross, at \(\mathcal A = 6, \xi = -1\).}
		\label{fig:score}
	\end{figure*}
	The maximum score is obtained for ($\mathcal{A} = 6,\xi = -1$), its location on the diagram is marked with a purple cross. A typical size for a crista is $R\approx$\qty{100}{nm},\cite{davies2011} which gives $C_0\approx$-\qty{0.01}{\per nm}, in agreement with recent simulations.\cite{konar2023} The pressure difference $P$ is actually not know, but we expect it to be much smaller than atmospheric pressure, which is in agreement with the value of $\mathcal{A}$ reported here. 
	The corresponding crista is shown as a 3D shape in Fig.~\ref{fig:vesicle}.
We see that under the influence of the proton field there is a clear partition of the curvature values with a flat membrane at the pole and a highly curved membrane at the equator.

	\begin{figure}
		\centering
		\includegraphics[scale=.35]{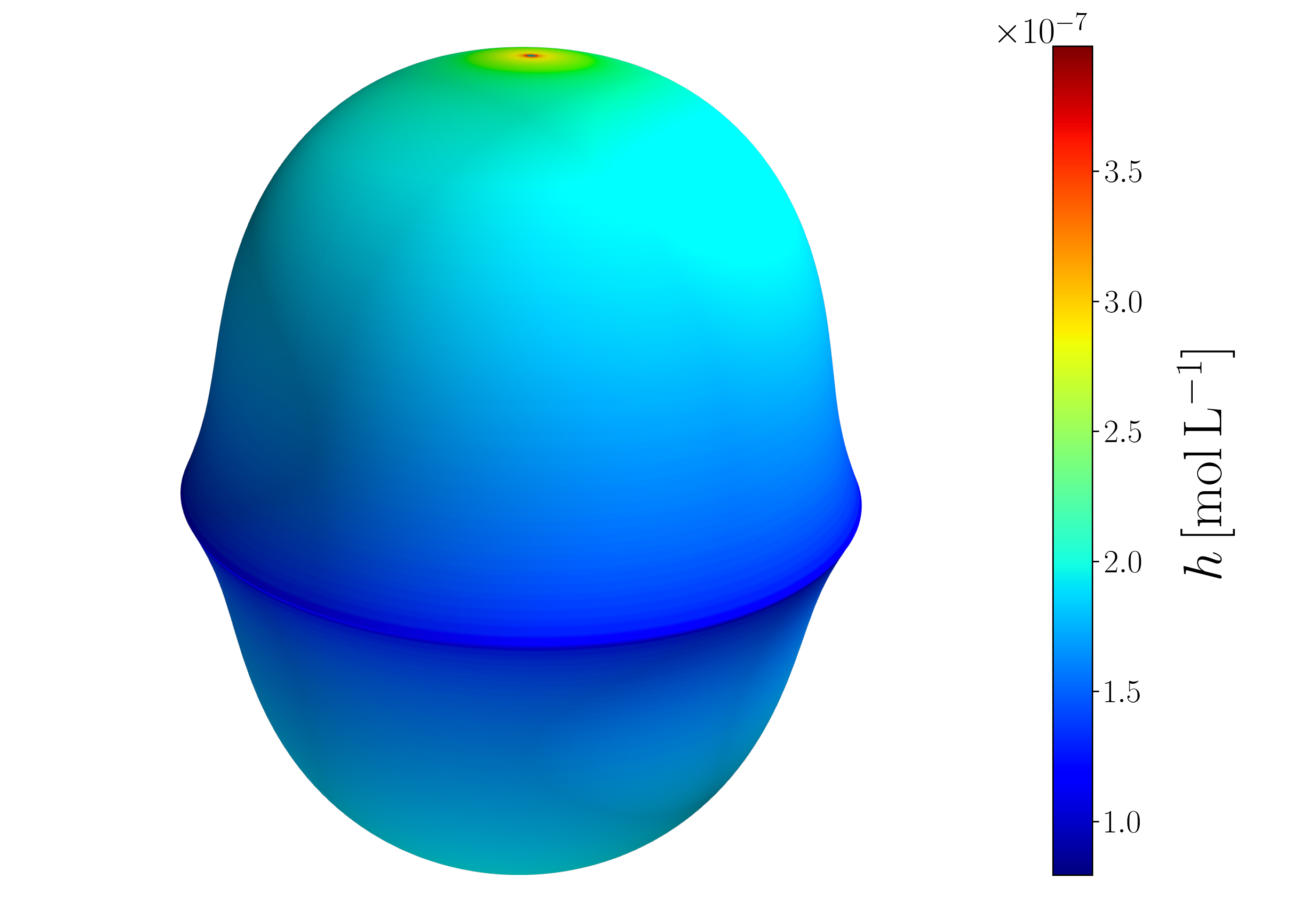}
		\caption{3D shape of an active vesicle with optimal functionality score, for \(\mathcal A = 6\) and \(\xi = -1\). The color code corresponds to the value of $h$ referenced on the color bar.}
		\label{fig:vesicle}
	\end{figure}
	
	\section{Discussion and conclusion}
	Active surfaces are surfaces whose mechanical properties are regulated by chemical species.\cite{Salbreux2017} Such models have been introduced by biophysicists to describe the mechanical properties of the cortex of cells, epithelial sheets,\cite{mietke,berthoumieux2014} etc. with great success. For these systems, the characteristic length scale is at least one micrometer and the chemical composition of the system is treated at the coarse-grained level.
	Other surfaces play a fundamental role in the cell: the lipid membranes that surround the organelles. Their properties and shapes also result from the coupling between the mechanical properties of the membranes and out-of-equilibrium chemical processes.
	The scale for the system is much smaller than for living cells or tissues as the typical size of a crista being tens of nanometres. It leads to a description at the molecular level and to a more precise description of the chemical processes that are involved. \par 
	Here, we model the crista membrane as an active membrane whose shape is driven by the surface proton concentration. We describe a crista as a closed vesicle modeled with a pH-dependent Helfrich Hamiltonian
	submitted to a surface diffusive flux of protons generated from RC to ATP-S. We show that such a system can adopt shapes with inhomogeneous curvature. In the relevant parameter space, we identify the zone leading to both highly curved in the ATP-S location and flat zones in the RC location. These shapes correspond to a well-functioning crista.\par
	We have shown that the diffusion of protons can shape the crista to become functional, using a simple pH sensitivity of the membrane, based on homogeneous CL concentration and without feedback with the deformation. The reality is more complex, as CL is thought to accumulate in curved zones.\cite{boyd2017,beltran2019} Once the membrane is deformed, its local chemical composition changes under this effect and so does its pH sensitivity. 
	The coupling between lipid composition and driving force, {\it i.e.} proton centration, is also a key point in understanding the function of cristae. Protons are trapped in interfaces for entropic reasons, whatever the interface is but the characteristic of their surface transport along lipid membranes depend on the lipid composition of the membrane.\cite{variyam24} The coupling between proton flux and chemical composition is still elusive but is a central point be better understood to propose a complete scheme of a working crista.\par
	In further work, a more realistic proton flux/chemical composition/mechanical properties coupling for the crista membrane needs to be investigated. 

	\begin{widetext}
		\newpage
		\vspace{3em}
		\appendix
		\begin{center}
			 \large \textbf 
			 {APPENDICES}
		\end{center}
		\section{Differencial geometry for a deformed sphere}
		We consider a deformed sphere $\mathbf{X}$ defined in Eq.~(\ref{deformed_sphere}). We derive the elements of differential geometry associated with this surface at the first order in the deformation field $u$, in the form $f+\delta f$, where $f$ is the zeroth order term associated with the perfect sphere and $\delta f$ the first order correction.
		\subsection{Differencial elements for the perfect sphere} \label{ap:geodiff0}
		We use standard notations for parametrizing the sphere, with $\theta$ the polar angle and $\phi$ the azimutal angle. ($\hat{\mathbf{x}}_\theta$,$\hat{\mathbf{x}}_\phi$,$\hat{\mathbf{x}}_r$) notes the spherical basis.
		The intrinsic basis of the perfect sphere $\mathbf{X}_0$ is defined as follow
		\begin{align}
			\mathbf{e}_\theta=\partial_\theta \mathbf{X}_0=R\hat{\mathbf{x}}_\theta,\quad 	\mathbf{e}_\phi=\partial_\phi \mathbf{X}_0 =R\sin(\theta)\hat{\mathbf{x}}_\phi
		\end{align}
		Thus the metric, $g_{ab}=\mathbf{e}_a\mathbf{e}_b$, and its inverse are given by
		\begin{align}
			g_{ab}=\begin{pmatrix}
				R^2 & 0\\
				0 & R^2\sin(\theta)^2
			\end{pmatrix},\quad 
			g^{ab}=\begin{pmatrix}
				R^{-2} & 0\\
				0 & R^{-2}\sin(\theta)^{-2}
			\end{pmatrix}, 
		\end{align}
		and the determinant of the $g_{ab}$, $g=R^4\sin(\theta)^2$.
		Finally, the Christoffel symbols of the Levi-Civita connection are:
		
		\begin{gather}
			\Gamma_{ab\theta}=R^2\begin{pmatrix}
				0 & 0\\
				0 & -\sin(\theta)\cos(\theta)
			\end{pmatrix},\quad 
			\Gamma_{ab\phi}=R^2\begin{pmatrix}
				0 & \sin(\theta)\cos(\theta)\\
				\sin(\theta)\cos(\theta) & 0
			\end{pmatrix}, \\ \Gamma_{ab}^\theta=\begin{pmatrix}
				0 & 0\\
				0 & -\sin(\theta)\cos(\theta)
			\end{pmatrix},\quad 
			\Gamma_{ab}^\phi=\begin{pmatrix}
				0 & {\rm cotan}(\theta)\\
				{\rm cotan}(\theta) & 0
			\end{pmatrix}
		\end{gather}
		
		The normal vector $\mathbf{n}$ is $\hat{\mathbf{x}}_r$ and the curvature tensor $K_{ab}=\mathbf{e}_a\cdot\partial_b\mathbf{n}$ is:
		\begin{align}
			K_{ab}=R\begin{pmatrix}
				1 & 0\\
				0 & \sin^2(\theta)
			\end{pmatrix}
		\end{align}
		and the mean and Gaussian curvature obey:
		\begin{align}
			C=K_a^a=2/R, \quad K_G=det(K_a^b)=R^{-2}.
		\end{align}
		\subsection{Correction for the deformed sphere} \label{ap:geodiff1}
		The first order correction of the intrinsic basis is
		\begin{equation} 
			\delta \mathbf{e}_\theta = \partial_\theta \mathbf{X}_1= Ru\hat{\mathbf x}_\theta+Ru'\hat{\mathbf x}_\phi \quad \delta \mathbf{e}_\phi= \partial_\phi \mathbf{X}_1=Ru\sin(\theta)\hat{\mathbf x}_\phi
		\end{equation}
		Thus, the first correction for the metric and its inverse is written as follows : 
		\begin{align}
			\delta g_{ab} = R^2\begin{pmatrix}
				2u&0\\
				0&2\sin^2\theta \, u\\
			\end{pmatrix}, \quad
			\delta g^{ab} = R^{-2}\begin{pmatrix}
				-2u&0\\
				0&-2\sin^{-2}\theta \, u\\
			\end{pmatrix}
		\end{align}; and \(\delta g = 4gu\).
		~\\~\\
		Finally, the Christoffel symbols of the Levi-Civita connection are corrected by the first order term:
		\begin{gather}
			\delta\Gamma_{ab\theta} =R^2\begin{pmatrix}
				u'&0\\
				0&-\sin2\theta u - \sin^2\theta u'\\
			\end{pmatrix}, \quad
			\delta\Gamma_{ab\phi} =R^2\begin{pmatrix}
				0&\sin2\theta u + \sin^2\theta u'\\
				\sin2\theta u + \sin^2\theta u'&0\\
			\end{pmatrix}, 
			\\
			\delta\Gamma_{ab}^\theta =\begin{pmatrix}
				u'&0\\
				0&-\sin^2\theta\ u'\\
			\end{pmatrix}, \quad
			\delta\Gamma_{ab}^\phi = \begin{pmatrix}
				0&u'\\
				u'&0\\
			\end{pmatrix}
		\end{gather} 
		
		The normal vector is corrected by the first order term : $\delta\mathbf n = -u'\, \hat{\mathbf x}_\theta$; 
		The first order correction for the curvature tensor is:\begin{align}
			\delta K_{ab} = R\begin{pmatrix}
				u-u''&0\\
				0&\sin^2\theta u - \sin\theta\cos\theta u'\\
			\end{pmatrix}
		\end{align} 
		We obtain the correction for the mean and of the Gaussian curvature: 
		\begin{align}
			\delta C = -R^{-1}\left(2u + \cot\theta u' + u''\right), \quad \delta K_G = -R^{-2}\left(2u + \cot\theta u' + u''\right).
		\end{align}

	\section{Solution for $\mathcal{A}$=0}\label{ap:a=0}
		For practical purposes, in Appendices~\ref{ap:a=0},\ref{ap:a=2}, we introduce the notation \(t(\theta)\) as: \(t(\theta) = \tan\sfrac{\theta}{2}\), and we implicitly evaluate trigonometric functions at the angle \(\theta\).
		
		In the case where \(\mathcal A = 0\), we no longer have a particular solution to Eq.~(\ref{deformation}) of the form given in Eq.~(\ref{part_sol}). We need to solve \(\laplace2\Delta_{\mathcal{S}} u_{\rm p} = 2(1-\xi)\beta\alpha_0C_{0,1} + (2(1-\xi)\beta\alpha_1 + \delta p)C_{0,0}\). Let \(f_{0,0}, f_{0,1}\) be two functions such that \(\Delta_{\mathcal{S}} f_{0,0} = C_{0,0},\; \Delta_{\mathcal{S}} f_{0,1} = C_{0,1}\). Then, a particular solution to Eq.~(\ref{part_sol}) is: \( u_{\rm p} = (1-\xi)\beta\alpha_0f_{0,1} + \left((1-\xi)\beta\alpha_1+\sfrac12\delta p\right)f_{0,0} \).
		\subsection{\(\Delta_{\mathcal{S}} f_{0,0} = 1\)}
		We determine \(f'_{0,0}\), ruled by a first order equation, then we integrate it:
		\begin{equation}
			f'_{0,0} = -\cot, \quad f_{0,0} = -\ln(\sin)
		\end{equation}
		\subsection{\(\Delta_{\mathcal{S}} f_{0,1} = -\ln t\)}
		We integrate by parts the first order equation on \(f'_{0,1}\), then we use Spence's function to solve \(f_{0,1}\) (see Eqs.~\ref{eq:dilog1}, \ref{eq:dilog2})
		\begin{equation}
			 f_{0,1}' = \cot\ln t - \frac{\ln\sin}{\sin}, \quad f_{0,1} = \big[\ln\sin\ln t\big] - \int 2\frac{\ln\sin}{\sin} = \ln\sin\ln t - \ln^2(2t) - \Li(-t^2)
		\end{equation}
		\subsection{Particular solution for \(\mathcal A = 0\)}
		A particular solution is then:
		\begin{equation}
			 u_{\rm p} = \left(1-\xi\right)\beta\alpha_0\left(\ln\sin\ln t - \ln^2(2t) - \Li(-t^2)\right) - \left(\left(1-\xi\right)\beta\alpha_1+\sfrac{\delta p}{2}\right)\ln\sin
		\end{equation}
	
	\section{Solution for $\mathcal{A}$=2} \label{ap:a=2}
		In the case where \(\mathcal A = 2\), \(C_{\mathcal A, b} = C_{2,b}\), and we no longer know the kernel of the deformation equation Eq.~(\ref{deformation}). We know that \(C_{2,0},C_{2,1}\) are in it, thus we are missing two functions. Let \(f_{2,0},f_{2,1}\) such that \(\laplace2f_{2,b} = C_{2,b}\). They are obviously linearly independent with \(C_{2,0}, C_{2,1}\) and with each other; by definition \(\laplace2^2f_{2,b} = 0\). Thus, the kernel is generated by the family \(\left(C_{2,0}, C_{2,1}, f_{2,0}, f_{2,1}\right)\). The expression for \(C_{2,1} = \cos\) is known; in order to determine \(C_{2,0}, f_{2,0}, f_{2,1}\), we develop a generic method for the following problem for any function \(b\):
		\begin{equation}
			\label{eq:genmeth2}
			\laplace2f = b
		\end{equation}
	Using variation of parameters and integration by parts, we introduce \(g,k,\lambda\) such that:
		\begin{gather}
		 \notag f = g\cos,\quad g= \int\lambda k,\quad\lambda = \int \cos\sin b,\quad k = \cos^{-2}\sin^{-1}\\
		\mathrm{i.e.}\quad f = \cos\int\left(\frac{1}{\cos^2\sin}\int \cos\sin b\right) 
		\end{gather}
		\subsection{\(\laplace 2 f = 0\)}
		We look for a solution of the form \(f = g\cos\), \(g = \int \lambda \), \(\lambda' = 0\). Evidently, \(\lambda\) is a constant and we choose \(g' = k\):
		
		\begin{equation}
			g = \ln t + \frac{1}{\cos}, \quad f = g\cos = \cos \ln t+ 1
		\end{equation}
		
		This solution is even, it is thus coherent to choose \(C_{2,0} (\theta)= \cos(\theta)\ln\tan(\sfrac\theta2) + 1\).
		\subsection{\(\laplace2f_{2,0} = \cos\ln t+ 1\)}
		We look for a solution in the form \(f = g\cos\), \(g = \int \lambda k\), \(\lambda' = \cos\sin \left(\cos\ln t + 1\right)\).
		
		\begin{gather}
			\notag \lambda = -\frac{\cos^3}{3}\left(\ln t + 1\right) + \frac{\ln\sin}{3}, \quad g = \sfrac13\left[-\ln t - \ln\sin - \ln\sin\ln t + \frac{\ln\sin}{\cos} + \ln^22t + \Li\left(-t^2\right)\right],\\
			f_{2,0}(\theta) = \sfrac13\left[-\cos\ln t - \cos\ln\sin - \cos\ln\sin\ln t + \ln\sin + \cos\ln^22t + \cos\Li\left(-t\right)\right]
		\end{gather}
		where we used Eqs.~\ref{eq:dilog1},\ref{eq:dilog2} to solve g.
		\subsection{\(\laplace2f_{2,1} = \cos\)}
		We look for a solution in the form \(f_{2,1} = g\cos\), \(g = \int \lambda k\), \(\lambda' = \cos^2\sin\).
		The equation on \(\lambda\) is: 
		\begin{equation}
			\lambda = -\frac{\cos^3}{3}, \quad g = -\sfrac13\ln\sin,\quad f_{2,1} = -\sfrac13\cos\ln\sin
		\end{equation}
		 
		\subsection*{How to compute \(\int 2\frac{\ln\sin}{\sin}\)}
		Using the tangent half-angle formula \(\sin = \frac{2t}{1+t^2}\), we change variables: \(\tau = t(x)\), \(\mathrm d\tau =\frac{1+t^2(x)}{2}\mathrm dx\).
		\begin{align}
			\begin{split}
				\label{eq:dilog1}
				\int^\theta 2\frac{\ln\sin}{\sin} &= \int^\theta 2\frac{\ln(2t(x)) - \ln(1+t^2(x))}{t(x)} \frac{1+t^2(x)}{2}\mathrm dx\\
				&= \int^{t(x)} 2\frac{\ln(2\tau) - \ln(1+\tau)}{\tau} \mathrm d\tau\\
				&= \ln^2(2t(x)) - \int^{t(x)} 2\frac{\ln(1+\tau^2)}{\tau}\mathrm d\tau
			\end{split}
		\end{align}
		Changing again variables: \(u = -\tau^2\), \(\mathrm du = -2\tau\mathrm d\tau\). We get:
		\begin{align}
			\begin{split}
				\label{eq:dilog2}
				\int^\theta 2\frac{\ln\sin}{\sin}&= \ln^2(2t(\theta)) - \int^{-t^2(x)} \frac{\ln(1-u)}{\sqrt{-u}}\frac{-\mathrm du}{\sqrt{-u}}\\
				&= \ln^2(2t(\theta)) - \int^{-t^2(x)} \frac{\ln(1-u)}{u}\mathrm du\\
				&= \ln^2(2t(\theta)) + \Li(-t^2(\theta))
			\end{split}
		\end{align}
		where \(\Li\) is defined as the dilogarithm (or Spence's) function, \(\Li(x) = -\int^x_0 \frac{\ln\left(1-s\right)}{s}{\rm d}s\).
		
			\section{Expression of the matrix $\mathbf{M}$} \label{ap:matrix}
		We give here the expression of the vector \(\mathbf V\) and the matrix \(\mathbf M\) introduced in Eq.~\ref{lambda_solve}.
		
		\begin{equation}
			\mathbf V = \begin{pmatrix*}\arcsin (\theta_0),& \left(\xi - 1\right) \beta h(\theta_0),& -1, & 0, & 0, & \beta\left(1-\xi\right)\alpha_0/\mathcal A \end{pmatrix*}
		\end{equation}
		
		To simplify the expression of \(\mathbf M\), we define the vector:
		\begin{equation}
			c(\theta)=\Big(C_{2,0}(\theta), C_{2,1}(\theta), C_{\mathcal{A},0}(\theta),C_{\mathcal{A},1}(\theta),C_{0,0}(\theta),C_{0,1}(\theta)\Big)
		\end{equation}
		
		\begin{eqnarray}
			C_{z0}=c(\theta_0),\quad 	C_{p0}=c'(\theta_0),\quad 	C_{s0}=c''(\theta_0)
		\end{eqnarray}
		Using these notations, the matrix \(\mathbf M\) is equal to:

		\begin{equation}
			\resizebox{.9\hsize}{!}{$\begin{pmatrix} \displaystyle -r/RC_{z0}+\cos(\theta_0)C_{p0} \\(\xi -2\mathcal{A}-\Gamma_0\sin(\theta_0))C_{z0}+(1-\xi)\cot(\theta_0)C_{p0}-C_{s0}\\\begin{matrix}1 & \cot(\gamma/2) & 1 & \cot(\gamma/2) & 1 & \cot(\gamma/2) \\\xi -2\mathcal{A} + 2 -k\sin(\gamma/2) & 0 & \xi-\mathcal{A}-k\sin(\gamma/2) & 0 & \xi-2\mathcal{A}-k\sin(\gamma/2) & 0\\\frac{\sin(\theta_0)}{2}C'_{2,0}(\theta_0) & \frac{\sin(\theta_0)}{2}C'_{2,1}(\theta_0) & \frac{\sin(\theta_0)}{\mathcal{A}}C'_{\mathcal{A},0}(\theta_0) & \frac{\sin(\theta_0)}{\mathcal{A}}C'_{\mathcal{A},1}(\theta_0) &\cos(\theta_0) & \ln(\sin(\theta))-\cos(\theta_0)\ln(\tan(\frac{\theta_0}{2})) \\0 & 0 & 0 & 0 & 0 & 1 \end{matrix}\end{pmatrix}$}
		\end{equation}
		
		\section{Criteria for the truncation of the infinite series of \(C_{\mathcal A, b}\) in the general case} \label{ap:truncation}
			In the general case, when $\mathcal{A}$ is not of the form \(n[n+1], n \in \mathbb N\), the expressions for \(C_{\mathcal A, b}\) Eq.~(\ref{GF1}) don't simplify to a short analytical formula. Thus, we need to approximate both functions as well as their first derivatives to their \(N(\mathcal A)\)th truncation, with \(N(\mathcal A)\) big enough so that the truncation errors be arbitrarily small:
			{\everymath = {\displaystyle}
				\begin{equation}
					\left.\begin{array}{l}
						R_{N(\mathcal A), 0} (\theta) = \left| \sum_{n > N(\mathcal A)}\prod_{k = 0}^{n-1}\left[2k\left(2k+1\right)-\mathcal A\right] \frac{\cos^{2n}(\theta)} {\left(2n\right)!}\right| \vspace{.5em}\\
						R_{N(\mathcal A), 1}(\theta) = \left| \sum_{n > N(\mathcal A)}\prod_{k = 1}^{n}\left[\left(2k-1\right)2k-\mathcal A\right] \frac{\cos^{2n+1}(\theta)} {\left(2n+1\right)!}\right| \vspace{.5em}\\
						R'_{N(\mathcal A), 0} (\theta) = \left| \sum_{n > N(\mathcal A)}\prod_{k = 0}^{n-1}\left[2k\left(2k+1\right)-\mathcal A\right] \frac{\cos^{2n-1}(\theta)} {\left(2n-1\right)!}\right|\vspace{.5em}\\
						R'_{N(\mathcal A), 1}(\theta) = \left| \sum_{n > N(\mathcal A)}\prod_{k = 1}^{n}\left[\left(2k-1\right)2k-\mathcal A\right] \frac{\cos^{2n}(\theta)} {\left(2n\right)!}\right| 
					\end{array} \right\} \leqslant \epsilon \quad\forall \theta \in \left[\theta_0,\sfrac\pi2\right]
				\end{equation}
			}	
			These function are monotonous and diverge fast enough in 0 so that it suffices to verify these 4 conditions in \(\theta_0\); additionally we observe that \(R_{n, b} \leqslant R'_{n, b} \;\forall n, b\). Thus, we focus on the last two sums. In this appendix, we have detailed the solution for \(b = 0\), they can be adapted to apply to the case \(b = 1\). The final integer \(N(\mathcal A)\) is chosen as the maximum of the two solutions. 
			
			The subscript \(b\), here always equal to \(0\), and the evaluation of \(R_n\) in \(\theta_0\), are henceforth omitted. Let \(n\in\mathbb N\). We introduce the sequence \((b_m)\) defined as: 
			\begin{equation}
				b_m = -\mathcal A \prod_{k = 1}^{m-1}\frac{2k\left(2k+1\right)-\mathcal A}{2k\left(2k+1\right)} \mbox{ so that } R'_n = \left|\sum_{m>n}b_m\cos^{2m-1}(\theta_0)\right|
			\end{equation}
			The objective is to bound \(R_n\) from above with the sum of a geometrical series. We must distinguish two cases: \(\mathcal A < 0\) and \(\mathcal A > 0\) (exact solutions are known when \(\mathcal A = 0\)). 
			\subsection{\(\mathcal A > 0\)}
			For \(n\in\mathbb N, n \geqslant m_0 = \lfloor\sqrt{\sfrac{1}{16} + \sfrac{\mathcal A}{8}}\rfloor + 1, \left|b_{n+1}\right| < \left|b_{m_0}\right|\):
			\begin{equation}
				R_n \leqslant\left|b_{m_0}\right|\sum_{m>n} \cos^{2m-1}(\theta_0) = \left|b_{m_0}\right| \frac{\cos^{2n-1}(\theta_0)}{1 - \cos^2(\theta_0)}
			\end{equation}
			We want \(N(\mathcal A)\) such that \(\left|b_{m_0}\right| \frac{\cos^{2N(\mathcal A)-1}(\theta_0)}{1 - \cos^2(\theta_0)} < \epsilon\), we choose :
			\begin{equation}
				N(\mathcal A) = \max\left\{0,\lfloor\sfrac12 \left(\left[\ln\epsilon + \ln\left(1-\cos^2(\theta_0)\right) - \ln\left|b_{m_0}\right|\right]/\cos(\theta_0)+ 1\right)\rfloor\right\} + 1
			\end{equation}
			\subsection{\(\mathcal A < 0\)}
			Using the same notations as in the previous section, we clearly have: \(b_{m+1}>b_m > 0 \; \forall m\in\mathbb N\). However, the sequence\( \left(\frac{b_{m+1}}{b_m}\right)_{m \in \mathbb N}\) is descending and approaches 1. In particular, for \(m_0 = 1 +\lfloor\sfrac12\sqrt{-\mathcal A}\cot(\theta_0)\rfloor, \frac{b_{m_0+1}}{b_{m_0}}\cos^2\left(\theta_0\right) < 1\):
			\begin{equation}
				R_n \leqslant b_{m_0} \cos^{2n -1}(\theta_0)\sum_{m>0}\frac{b_{m_0+1}}{b_{m_0}}\cos^{2}(\theta_0)^m
				= \frac{b_{m_0} \cos^{2n -1}}{1 - \sfrac{b_{m_0+1}}{b_{m_0}}\cos^{2}(\theta_0)}
			\end{equation}
			We want \(N(\mathcal A)\) such that \(\frac{b_{m_0} \cos^{2n -1}}{1 - \sfrac{b_{m_0+1}}{b_{m_0}}\cos^{2}(\theta_0)} < \epsilon\), we choose :
			\begin{equation*}
				N(\mathcal A) = \max\left\{0,\lfloor\sfrac12 \left(\left[\ln\epsilon + \ln\left(1-\sfrac{b_{m_0+1}}{b_{m_0}}\cos^2(\theta_0)\right) - \ln b_{m_0}\right]/\cos(\theta_0)+ 1\right)\rfloor\right\} + 1
			\end{equation*}
	\end{widetext}
	\bibliographystyle{unsrt}
	\bibliography{biblio}

\begin{thebibliography}{10}

\bibitem{panek2020}
T.~P{\'a}nek, M.~Eli{\'a}s, M.~Vancov{\'a}, J.~Luke\v{s}, and H.~Hashimi.
\newblock Returning to the fold for lessons in mitochondrial crista diversity
  and evolution.
\newblock {\em Current Biology}, 30:575--588, 2020.

\bibitem{davies2011}
K.~M. Davies, M.~Strauss, B.~Daum, J.~H. Heinz, D.~Osiewacz, A.~Rycovska,
  V.~Zickermann, and W.~K{\"u}hlbrandt.
\newblock Macromolecular organisation of atp synthase and complex i in whole
  mitochondria.
\newblock {\em Proc. Nat. Acad. of Sci.}, 108:14121--14126, 2011.

\bibitem{serowy2003}
S.~Serowy, S.~M. Saparov, Y.~N. Antonenko, W.~Kozlovsky, V.~Hagen, and P.~Pohl.
\newblock Structural proton diffusion along lipid bilayers.
\newblock {\em Biophysical Journal}, 84:1031--1037, 2003.

\bibitem{knyazev}
D.~G. Knyazev, T.~P. Silverstein, S.~Brescia, A.~Maznichenko, and P.~Pohl.
\newblock A new theory about interfacial proton diffusion revisited: The
  commonly accepted laws of electrostatics and diffusion prevail.
\newblock {\em Biomolecules}, 13:1641, 2023.

\bibitem{heberle1994}
J.~Heberle, G.~Riesle, J.~Thiedemann, D.~Oesterhelt, and N.~A. Dencher.
\newblock Proton migration along the membrane surface and retarded surface to
  bulk transfer.
\newblock {\em Nature}, 370, 1994.

\bibitem{heberle2000}
J.~Heberle.
\newblock Proton transfer reactions across bacteriorhodopsin and along the
  membrane.
\newblock {\em BBA-Bioenergetics}, 1458, 2000.

\bibitem{gennis2016}
R.~B. Gennis.
\newblock Proton dynamics at the membrane surface.
\newblock {\em Biophysical Journal}, 110, 2016.

\bibitem{cherepanov2003low}
D.~A. Cherepanov, B.~Feniouk, , W.~Junge, and Mulkidjanian~A. Y.
\newblock Low dielectric permittivity of water at the membrane interface:
  effect on the energy coupling mechanism in biological membranes.
\newblock {\em Biophysical journal}, 85:1307--1316, 2003.

\bibitem{paradies2014}
G.~Paradies, V.~Paradies, V.~De~Benedictis, F.~M. Ruggiero, and G.~Petrosillo.
\newblock Functional role of cardiolipin in mitochondrial bioenergetics.
\newblock {\em BBA}, 1837:408--417, 2014.

\bibitem{ikon2017}
N.~Ikon and R.~O.. Ryan.
\newblock Cardiolipin and mitochondrial cristae organization.
\newblock {\em BBA}, 1859:1156--1163, 2017.

\bibitem{beltran2019}
E.~Beltrán-Heredia, F-C. Tsai, S.~Salinas-Almaguer, F.~J. Cao, P.~Bassereau,
  and F.~Monroy.
\newblock Membrane curvature induces cardiolipin sorting.
\newblock {\em Comm. Biol}, 2:225, 2019.

\bibitem{haines2002}
N.~A.~Dencher T.~H.~Haines.
\newblock Cardiolipin: a proton trap for oxidative phosphorylation.
\newblock {\em FEBS}, 528, 2002.

\bibitem{variyam24}
A.~R. Variyam, M.~Rzycki, A.~Yucknovsky, A.~A. Stuchebrukhov, and N.~Drabik,
  D.~Amdursky.
\newblock Proton diffusion on the surface of mixed lipid membranes highlights
  the role of membrane composition.
\newblock {\em Biophysical Journal}, 123:1--11, 2024.

\bibitem{khalifat2008membrane}
N.~Khalifat, N.~Puff, S.~Bonneau, J.-B. Fournier, and M.~I. Angelova.
\newblock Membrane deformation under local ph gradient: mimicking mitochondrial
  cristae dynamics.
\newblock {\em Biophysical journal}, 95:4924--4933, 2008.

\bibitem{khalifa2011}
N.~Khalifat, J-B. Fournier, M.~I. Angelova, and N.~Puff.
\newblock Lipid packing variations induced by ph in cardiolipin-containing
  bilayers: The driving force for the cristae-like shape instability.
\newblock {\em BBA-Biomenbranes}, 1808:2724--2723, 2011.

\bibitem{allolioc2021}
C.~Allolio and D.~Harries.
\newblock Calcium ions promote membrane fusion by forming negative curvature
  inducing clusters on specific anionic lipids.
\newblock {\em ACS nano}, 15:12880--12887, 2021.

\bibitem{joubert2021}
F.Joubert and N.~Puff.
\newblock Mitochondrial cristae architecture and functions: lessons from
  minimal model systems.
\newblock {\em Membranes}, 11:11070465, 2021.

\bibitem{helfrich1973}
W.~Helfrich.
\newblock Elastic properties of lipid bilayers—theory and possible experi-
  ments.
\newblock {\em Z. Naturforsch.}, 33:305--315, 1973.

\bibitem{helfrich1989}
Ou-Yang Zhong-can and W.~Helfrich.
\newblock Bending energy of vesicle membranes: General expressions for the
  first, second and third variation of the shape energy and applications to
  spheres and cylinder.
\newblock {\em Phys. Rev. A}, 39:5280--5288, 1989.

\bibitem{seifert1997}
U.~Seifert.
\newblock Configurations of fluid membranes and vesicles.
\newblock {\em Advances in Physics}, 46:13--137, 1997.

\bibitem{mietke}
A.~Mietke, F.~J{\"u}licher, and I.~F. Sbalzarini.
\newblock Self-organized shape dynamics of active surfaces.
\newblock {\em Proc. Natl. Acad. Sci. U. S. A.}, 116:29--34, 2019.

\bibitem{thesemietke}
A.~Mietke.
\newblock {\em Dynamics of active surfaces}.
\newblock PhD thesis, TU Dresden, 2018.

\bibitem{Guv04a}
J.~Guven.
\newblock Membrane geometry with auxiliary variables and quadratic constraints.
\newblock {\em Journal of Physics A: Mathematical and General}, 37:L313, 2004.

\bibitem{Deserno_curvatureconvention}
M.~Deserno.
\newblock Fluid lipid membranes: From differential geometry to curvature
  stresses.
\newblock {\em Chemistry and physics of lipids}, 185:11--45, 2015.

\bibitem{AF2}
A-F. Bitbol, N.~Puff, Y.~Sakuma, M.~Imai, J-B. Fournier, and M.~I. Angelova.
\newblock Lipid membrane deformation in response to a local ph modification:
  theory and experiments.
\newblock {\em Soft Matter}, 8:6073--6082, 2012.

\bibitem{patil2020}
N.~Patil, S.~Bonneau, A-F.~Bitbol F.~Joubert, and H.~Berthoumieux.
\newblock Mitochondrial cristae modeled as an out-of-equilibrium membrane
  driven by a proton field.
\newblock {\em Phys. Rev. E}, 102:0224401, 2020.

\bibitem{mendes2023}
T.~V. Mendes, J.~Ranft, and H.~Berthoumieux.
\newblock Model of membrane deformations driven by a surface gradient.
\newblock {\em Phys. Rev. E}, 108:014113, 2023.

\bibitem{blum2019}
T.~B. Blum, A.~Hahn, T.~Meier, K.~M. Davies, and W.~K{\"u}hlbrandt.
\newblock Dimers of mitochondrial atp synthase induce membrane curvature and
  self-assemble into rows.
\newblock {\em Proc. Natl. Acad. Sci. U. S. A.}, 116, 2019.

\bibitem{rieger2014lateral}
B.~Rieger, W.~Junge, and K.~B Busch.
\newblock Lateral ph gradient between oxphos complex iv and f0f1 atp-synthase
  in folded mitochondrial membranes.
\newblock {\em Nature communications}, 5:3103, 2014.

\bibitem{Muller_phd}
M.~M. M{\"u}ller.
\newblock {\em Theoretical studies of fluid membrane mechanics}.
\newblock PhD thesis, Johannes Gutenberg-Universit{\"a}t Mainz, 2007.

\bibitem{fournier2007}
J.-B. Fournier.
\newblock On the stress and torque tensors in fluid membranes.
\newblock {\em Soft Matter}, 3:883--888, 2007.

\bibitem{helfrich1987}
O.~Zhong-can and W.~Helfrich.
\newblock Instability and deformation of a spherical vesicle by pressure.
\newblock {\em Phys. Rev. Lett.}, 59:5280--5288, 1987.

\bibitem{konar2023}
S.~Konar, H.~Arif, and C.~Allolio.
\newblock Mitochondrial membranes: model lipid compositions, material
  properties and the changing curvature of cardiolipin.
\newblock {\em Biophysical Journal}, 122:4274--4287, 2023.

\bibitem{sjoholm2017}
J.~Sj{\"o}holm, J.~Bergstrand, T.~Nilsson, R.~Šachl, C.~von Ballmoos,
  J.~Widengren, and P.~Brzezinski.
\newblock The lateral distance between a proton pump and atp synthase
  determines the atp-synthesis rate.
\newblock {\em Scientific Reports}, 7:2926, 2017.

\bibitem{Salbreux2017}
G.~Salbreux and F.~Jülicher.
\newblock Mechanics of active surfaces.
\newblock {\em Phys. rev. E}, 96:032404, 2017.

\bibitem{berthoumieux2014}
H.~Berthoumieux, J.-L. Maître, C.-P. Heisenberg, E.~K. Paluch,
  F.~J{\"u}licher, and G.~Salbreux.
\newblock Active elastic thin shell theory for cellular deformations.
\newblock {\em New Journal of Physics}, 16:065005, 2014.

\bibitem{boyd2017}
K.~J. Boyd, N.~N. Alder, and E.~R. May.
\newblock Buckling under pressure: Curvature-based lipid segregation and
  stability modulation in cardiolipin-containing bilaye.
\newblock {\em Langmuir}, 33:6937–6946., 2017.

\end{thebibliography}
\end{document}